\documentclass[twoside,slac_one]{revtex4}
\usepackage{graphicx}
\usepackage{fancyhdr}
\usepackage{amsmath} 
\usepackage{bm}
\usepackage{amsxtra}
\usepackage{amssymb}
\usepackage{amsthm}
\usepackage{latexsym}
\usepackage{lscape}

\newcommand{\MET}{\rlap{\kern0.25em/}E_T}
\newcommand{\pt}{$P_{\rm T}$}

\newcommand{\pythia}{{\sc{pythia}}}
\newcommand{\fewz}{{\sc{fewz}}}
\newcommand{\madgraph}{{\sc{madgraph}}}
\newcommand{\vbp}{{\sc{vbp}}}
\newcommand{\resbos}{{\sc{resbos}}}
\newcommand{\powheg}{{\sc{powheg}}}

\begin{document}

\title{The Angular Coefficients and $A_{fb}$ of Drell-Yan $e^{+}e^{-}$ Pairs in the Z Mass Region from $p\bar{p}$ Collision at $\sqrt{s}$ = 1.96 TeV}

%

\author{Jiyeon Han}
\affiliation{Department of Physics and Astronomy, University of Rochester, Rochester, NY, USA}

\begin{abstract}
We report on the measurement of angular coefficients and the forward and backward asymmetry ($A_{fb}$)
 of Drell-Yan dielectron pairs from $p\bar{p}$ collisions at $\sqrt{s}$ = 1.96 TeV.
The angular distributions are studied as a function of the transverse momentum of dielectron pair
 and $A_{fb}$ is measured using the event counting method.
The Lam-Tung ($A_{0}-A_{2}$ = 0) relation which is only valid for a spin-1 description of the gluon
 is also tested.
\end{abstract}

\maketitle

\thispagestyle{fancy}


\section{Introduction}
The general expression \cite{mirkes2} for  angular distribution of the final
state electron in
the  Collins-Soper(CS) frame \cite{collins} is given by:
\begin{eqnarray}
  \frac{d{\sigma}}{d\cos{\theta}d\phi} &\propto& (1+\cos^2{\theta})  \nonumber \\
  &+& \frac{1}{2}A_0(1-3\cos^2{\theta})
  + A_1\sin{2\theta}\cos{\phi} \nonumber \\
  &+&\frac{1}{2}A_2\sin^2{\theta}\cos{2\phi}
  + A_3\sin{\theta}\cos{\phi} \nonumber \\
  &+& A_4\cos{\theta} +A_5 \sin^2 \theta \sin 2\phi  \nonumber \\
  &+& A_6\sin{2\theta}\sin{\phi}
  + A_7\sin{\theta}\sin{\phi}.
  \label{AngleFunc}
\end{eqnarray}
Here,  $\theta$ and $\phi$ are the polar
and azimuthal angles of the electron in 
the CS frame.  
The angular coefficients, $A_0$ to $A_7$, are in general functions of the
invariant mass $M_{\ell\ell}$, rapidity $y$, and
transverse momentum $P_T$ of the dilepton  
in the lab frame \cite{cdf_coord}. 
When integrated over $\phi$, the differential cross section reduces to:
\begin{eqnarray}
  \frac{d{\sigma}}{d\cos{\theta}} \propto (1+\cos^2{\theta})
  + \frac{1}{2}A_0(1-3\cos^2{\theta}) + A_4\cos{\theta}
  \label{etheta}
\end{eqnarray}
When integrated over $cos\theta$, the differential cross section reduces to:
\begin{eqnarray}
  \frac{d{\sigma}}{d\phi} \propto 1+\beta_3\cos{\phi} + \beta_2 \cos 2{\phi} + \beta_7 \sin{\phi} + \beta_5 \sin 2{\phi}
  \label{ephi}
\end{eqnarray}

Calculations which are based on perturbative QCD (pQCD)  make definite predictions
for all of the angular coefficients.  For $p\bar{p}\to \gamma^{*}/Z\to e^{+}e^{-}~X$,
the angular coefficients  $A_5$, $A_6$
and $A_7$ are close to zero \cite{mirkes2}, and if we integrate over positive and negative y, 
the angular coefficients $A_1$ and $A_3$ are 
small.   We can use Eq. \ref{etheta} to extract $A_0$ and $A_4$ (averaged over $y$) and 
Eq. \ref{ephi} to extract $A_2$ and $A_3$ (averaged over $y$) under the assumption that 
$A_5$ and $A_7$ are zero (as is theoretically expected). 

The angular distribution of the final state electron
in the quark-antiquark ($q\bar{q}$) annihilation process  $q\bar{q}\to \gamma^{*}/Z\to e^{+}e^{-}$
(LO) can be written as
\begin{eqnarray}
  \frac{d\sigma}{dcos\theta} & \propto &  (1 + cos^{2} \theta) + B~cos\theta. 
  \label{qqbar}
\end{eqnarray}
Here  $A_{fb}(M_{\ell \ell})=\frac{3}{8}B$ is  the forward-backward asymmetry (which
originates from  the $\gamma^{*}/Z$ interference).

\begin{figure}[ht]
  \centering
  \includegraphics[width=70mm]{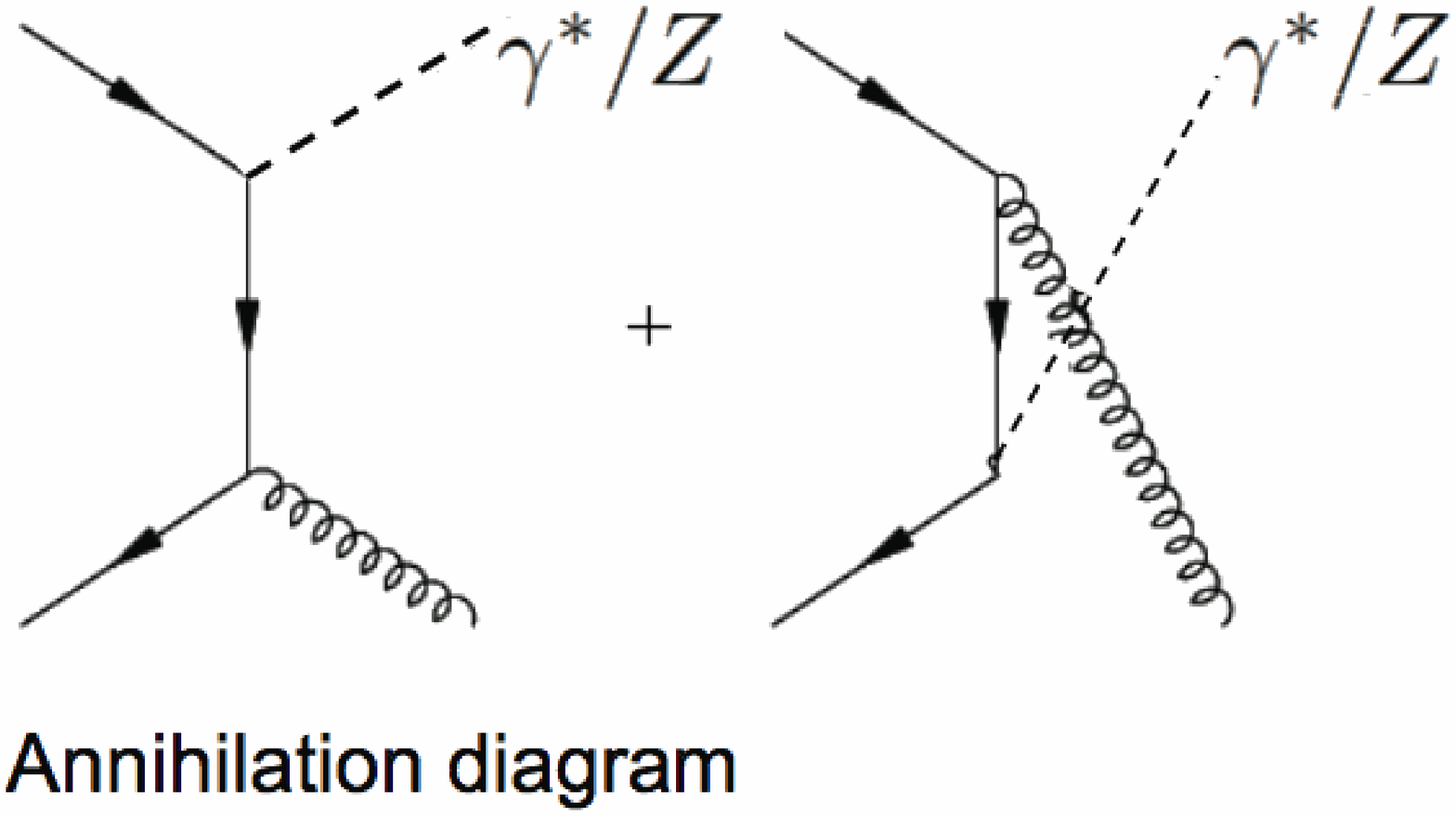}
  \includegraphics[width=70mm]{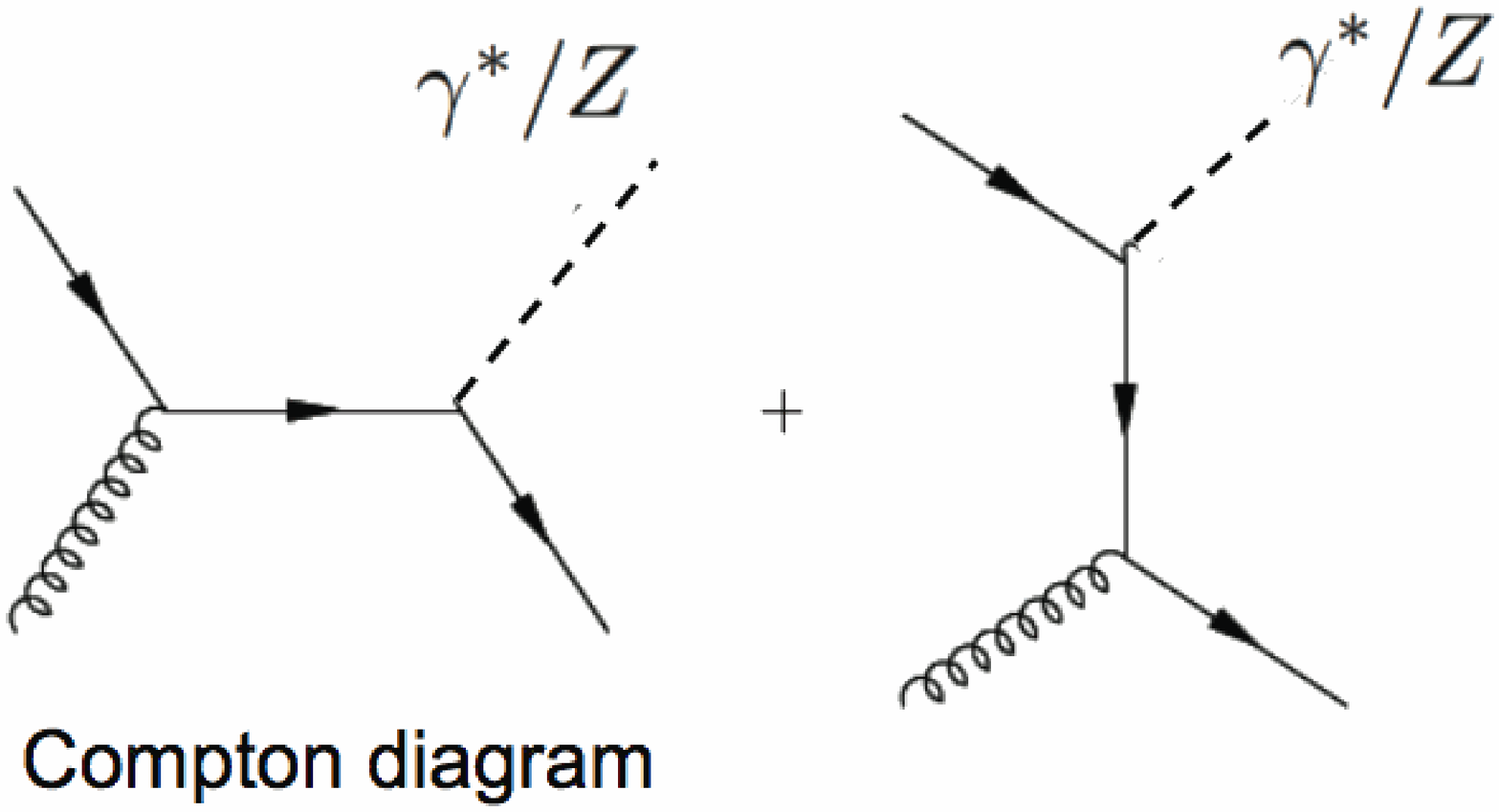}
  \caption{Leading order annihilation ($q\bar{q}\to \gamma^{*}/Z~g$) and Compton ($qg \to \gamma^{*}/Z~q$)
    diagrams for the production of $Z$ bosons with finite transverse momentum.} \label{fig:diagram}
\end{figure}

 In Quantum Chromodynamics (QCD) at the order of $\alpha_{s}$ (NLO)  this occurs
either through the annihilation process with  a gluon (g)  in the final state ($q\bar{q}\to \gamma^{*}/Z~g$),
or via  the Compton process with a quark in
the final state ($qg \to \gamma^{*}/Z~q$), as shown in figure \ref{fig:diagram}. 
 For the $q\bar{q}\to \gamma^{*}/Z~g$ annihilation process \cite{qq,bnl,berger,bodek},  
at NLO predicts that  the angular coefficients 
$A_0$ and $A_2$  are equal,
independent of  Parton Distribution Functions (PDFs),  or $y$,
and are described by  
$ A_0^{q\bar{q}} =  A_2^{q\bar{q}}= \frac{P_T^2}{M_{\ell\ell}^2+P_T^2}$.
For the $qg \to \gamma^{*}/Z~q$ Compton process, $A_0$ and $A_2$ 
depend on  PDFs and $y$.  However, at NLO, when averaged 
over y,  $A_0$ and $A_2$ are approximately \cite{gq,dyok}
described by 
 $A_0^{gq} = A_2^{qg} \approx \frac{5P_T^2}{M_{\ell\ell}^2+5P_T^2}$. 
The equality  $A_2= A_0$ is known as the Lam-Tung relation \cite{LT}.  At LO, it is valid for 
both  $q\bar{q}$ and $gq$ processes \cite{bnl}. 
Fixed order perturbative QCD calculations \cite{mirkes2} at NLO, as well as QCD  resummation
 calculations \cite{berger}  to all orders indicate that  violations of the Lam-Tung relation are small. 
The Lam-Tung relation is only valid for vector (spin 1) gluons. 
It is badly broken for  scalar (spin 0)  gluons \cite{scalar}.
Therefore,  confirmation of the Lam-Tung relation is  a fundamental test of the vector
gluon nature of  QCD and is equivalent to a measurement of the spin of the gluon.
Here, we measure the angular coefficients, $A_0$, $A_2$, $A_3$, and $A_4$ as a function of $P_{\rm T}$ and
 the first test of the Lam-Tung relation at large dilepton mass and high transverse momentum.
This measurement provides a detailed test of the production mechanism of gauge boson with finite $P_{\rm T}$. 
In addition, we measure $A_{fb}$ ($A_{fb}(M_{\ell \ell})=\frac{3}{8}A_{4}$) in mass which is sensitive
 to weak mixing angle, $\sin^{2}\theta_{W}$.
In the end, we compare the measurement with \pythia~and various predictions
 which are QCD resummation calculations and QCD calculations in the fixed order perturbation theories.

\section{Data Sample and Event Selection}

The data sample consists of 2.1 fb$^{-1}$ collected by CDF II Detector at Fermilab \cite{cdfdet}
 during 2004-2007 for the measurement of the angular coefficients. 
Charged particle directions and momenta are measured by an open-cell drift chamber (COT),
 a silicon vertex detector (SVX), and an intermediate silicon layer in a 1.4 T magnetic field.
Projective-tower-geometry calorimeters and outer muon detectors enclose the magnetic tracking volume.
The coverage of COT tracking in pseudorapidity is $|\eta|<1.2$.
The energies and directions of electrons, photons, and jets are measured by two separate calorimeters:
 central ($|\eta|<1.1$) and plug ($1.1<|\eta|<3.6$). Each calorimeter has an electromagnetic compartment
 with a shower maximum detector followed by a hadronic compartment.

The events come from the inclusive single high $p_{T}$ central electron or the dielectron trigger.
The electron trigger efficiencies as a function of $E_T$ are measured using the data.
The overall trigger efficiencies are measured to be almost $100\%$. \cite{dsdy_public,dsdy}
The dielectron data sample consists of three different $e^+e^-$ pair topologies. The CC topology has both electrons
in the central calorimeter with $|\eta|<1.1$. The CP topology has one electron in the central calorimeter with
$|\eta|<1.1$, and another in the plug calorimeter with $1.2<|\eta|<2.8$, a good fiducial region requirement.
 The PP topology has two electrons in the plug calorimeter with $1.2<|\eta|<2.8$.
Events with at least one electron candidate with $E_{T}>25$ GeV for
 CC and PP events, and $E_{T}>20$ GeV for CP events, are selected. 
The second electron candidate is required to have $E_{T}>15$ GeV for CC, 
$E_{T}>25$ GeV for PP, and $E_{T}>20$ GeV for CP events.
The asymmetry $E_{T}$ cut in CC topology increases the acceptance in high $P_{\rm T}$ region.
For all topologies, the $p\bar{p}$ collision vertex along the proton direction (z-axis) is required 
to be within 60 cm of the center of the detector.  

For the CC topology, oppositely charged electron pairs are required. 
One electron has tight and another has loose central electron ID cuts with COT track requirement \cite{idcut}. 
For the CP topology, the central electron passes the tight central electron ID cuts  
and the plug electron passes the standard plug electron ID cuts with a silicon track requirement \cite{idcut}. 
For the PP topology, both electrons pass the standard plug electron ID cuts \cite{idcut}. 
In addition, the pseudo-rapidity must have the same sign and both electrons must have a silicon track for the PP topology.
The track requirement in both legs of all topologies minimizes the background contamination down to $\sim 0.5 \%$,
 especially $\gamma+jet$ background as well as dijet background.
After the selection cuts, we find about 140,000 events for all (CC+CP+PP) topology.

For $A_{fb}$ measurement, we use twice more data, 4.1 fb$^{-1}$, with the inclusive single high $p_{T}$ central
 electron trigger sample only.
The event selection is same with the measurement of the angular coefficients except that
 $E_{T}>25$ GeV for both legs of CC topology is required  and PP topology is excluded in the measurement.
The PP topology has a large charge fake rate, so it is not optimal for the $A_{fb}$ measurement.

\section{Simulation Sample}

The effect of the acceptance on the angular distributions is modeled using the default 
\pythia~\cite{pythia} Monte Carlo (MC) generator combined
 with a {\sc geant} \cite{geant} simulation of the CDF detector \cite{cdfdet}.
The \pythia~generator used at CDF has additional ad-hoc tuning \cite{pythia}
 in order to accurately represent the Z boson transverse momentum distribution measured in data.
Further tuning was introduced in order to ensure that the MC correctly describes the rapidity,
 as well as the correlations between rapidity and transverse momentum that are observed in the data.
To reconstruct the simulated events in the same way as data, the calorimetry
 energy scale and resolution, and event selection efficiencies (electron ID and tracking)
 are tuned using data. 

\section{Backgrounds}

Due to the two track requirement in the selection, the background level is very small in the sample.
The remaning background is measured using data and MC simulations and subtract to get the pure signal.
The background from QCD process (mainly, QCD dijet) is measured using data.
$\gamma$+jet process is also possible background for Drell-Yan process, but the track requirment on 
both legs rejects most of $\gamma$+jet background.
The QCD background is measured using the isolation extrapolation method and the mass spectrum fitting method
from data directly. \cite{dsdy}
The electron is very isolated object, so the isolation energy shape is a good variable to distinguish the electron
 from the jet object which has a broad and flat isolation energy shape.
In the isolation extrapolation method, we fit the isolation distribution for both signal and background contributions
 and then extrapolate the background from the high isolation tail into the signal region.
To measure the background in $P_{\rm T}$, the background shape versus $P_{\rm T}$ is obtained from the background sample (data)
and its overall normalization is set using the background rate measured by the isolation extrapolation and mass 
spectrum fitting method.

The other background source is the background from Electroweak process.
The background from Electroweak process is obtained using MC simulations.
The Electroweak background processes considered are WW, WZ, inclusive $t\bar{t}$, inclusive W+jets,
and Z$\to \tau\tau$ process.
We generate the MC simulations for each process and estimate the background
for corresponding integrated luminosity.

\begin{figure}[ht]
  \centering
  \includegraphics[width=90mm]{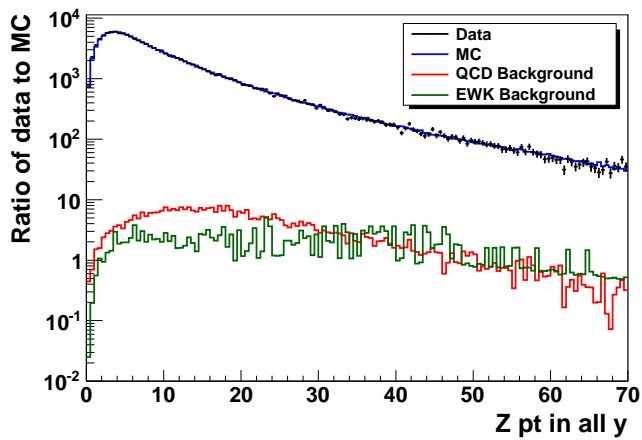}
  \caption{The $P_{\rm T}$ distribution of dielectron pairs.
    The event are selected in the full Z boson mass region, $66<M_{ee}<116$ GeV/$c^{2}$.
    The black crosses correspond to data, the blue histogram to CDF \pythia~MC.
    The red and green histogram correspond to QCD and Electroweak background, respectively.} \label{fig:zpt_spec}
\end{figure}

The overall QCD background is $0.30\%$ and the overall Electroweak background is $0.19\%$.
The background in $P_{\rm T}$ for QCD and Electroweak process is shown in  Figure \ref{fig:zpt_spec}.
All backgrounds are subtracted as a function of the angular distributions, $\cos\theta$ and $\phi$ 
for the angular coefficients measurement and mass for $A_{fb}$ measurement.

\section{Extraction of Angular Coefficients in $P_{\rm T}$}
The Collins-Soper \cite{collins} CM frame lepton decay angles, $\cos \theta$ and $\phi$, are defined by:

\begin{equation}
  \small
  \cos \theta = \frac{2}{M\sqrt{M^{2} + P_{\rm T}^{2}}} (\ell_{1}^{+}\ell_{2}^{-} - \ell_{1}^{-}\ell_{2}^{+})
  \label{CScost}
\end{equation} 

\begin{equation}
  \small
  \tan \phi = \frac{\sqrt{M^{2}+P_{\rm T}^{2}}}{M} \cdot \frac{\vec{\Delta}_{r}\cdot \hat{R}_{T}}{\vec{\Delta}_{r}\cdot \hat{Q}_{T}}
  \label{CSphi}
\end{equation} 
where $M$ is the mass of the dielectron pair, $\ell_{1}$ ($\ell_{2}$) is the 
four-momentum of electron (positron),
$P_{\rm T}$ is the transverse momentum of the electron-positron pair, 
$\ell^{\pm}$ corresponds to $\frac{1}{\sqrt{2}}(\ell^{0}\pm \ell^{3})$,
$\Delta^{j}=\ell^{j}-\bar{\ell}^{j}$, $\hat{Q_{T}}$ is a transverse unit vector in the direction of $\vec{Q_{T}}$,
and $\hat{R_{T}}$ is a transverse unit vector in the direction of $\vec{P_{A}}\times \vec{Q}$.

To extract the angular coefficients in $P_{\rm T}$, the data and the simulated MC (\pythia) are binned in five $P_{\rm T}$ bins which are
(0,10), (10,20), (20,35), (35,55), and (55 and above) using the reconstructed dielectron $P_{\rm T}$.
For each $P_{\rm T}$ range, the data and the MC simulated events are binned in $\cos\theta$ and $\phi$, respectively.
The MC events are re-weighted to generate the expected experimental distributions for a range of values of 
$A_0$ and $A_4$ for $\cos\theta$, and $A_2$ and $A_3$ for $\phi$ distribution.
Eq. \ref{etheta} and \ref{ephi} are used to re-weight the event to change the angular distributions
in the generated level.
A maximum log-likelihood comparison of the data to MC in $\cos\theta$ and $\phi$ is used to extract
the best values of the angular coefficients, ($A_0$,$A_4$) from $\cos\theta$ and ($A_2$,$A_3$) from $\phi$,
that describe the data.
To extract the angular coefficients in QED born level (before QED radiation), we reweight the events
using the angular coefficients obtained before QED radiation.

\subsection{Systematic Uncertainties}

\begin{figure}[ht]
  \centering
  \includegraphics[width=120mm]{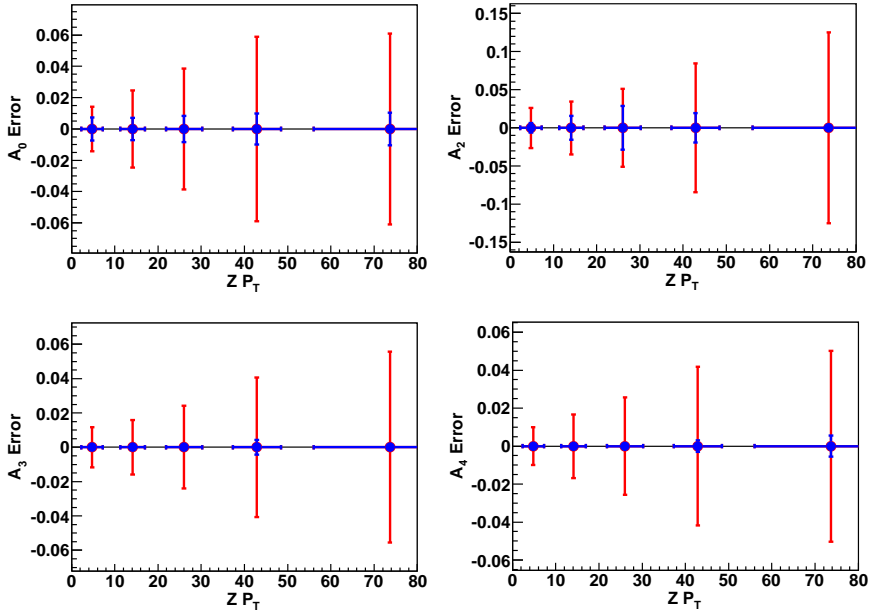}
  \caption{The systematic vs. statistical uncertainty for the angular coefficients.
    The red bars correspond to the statistical errors and the blue bars to the systematic errors.
    The systematic errors are very small compared to the statistical errors.} \label{fig:uncertainty}
\end{figure}

Systematic uncertainties are determined for the background estimation, the energy scale and resolution,
 the electron identification efficiency, the silicon tracking efficiency, the boson $P_{\rm T}$ and rapidity
 modeling, and material modeling.
The systematic uncertainties are considered in $\cos\theta$ and $\phi$ distribtuion for $P_{\rm T}$ bins.
An investigation of all of systematic errors listed above shows that the uncertainties on the extracted
 angular coefficients are dominated by the statistical errors.
 Figure \ref{fig:uncertainty} compares the level of systematic vs. statistical errors
 for the angular coefficients in $P_{\rm T}$.
For the total error, the statistical and systematic errors are added in quadrature.

\subsection{The Extracted Angular Coefficients in $P_{\rm T}$}
\begin{figure}[ht]
  \centering
  \includegraphics[width=80mm]{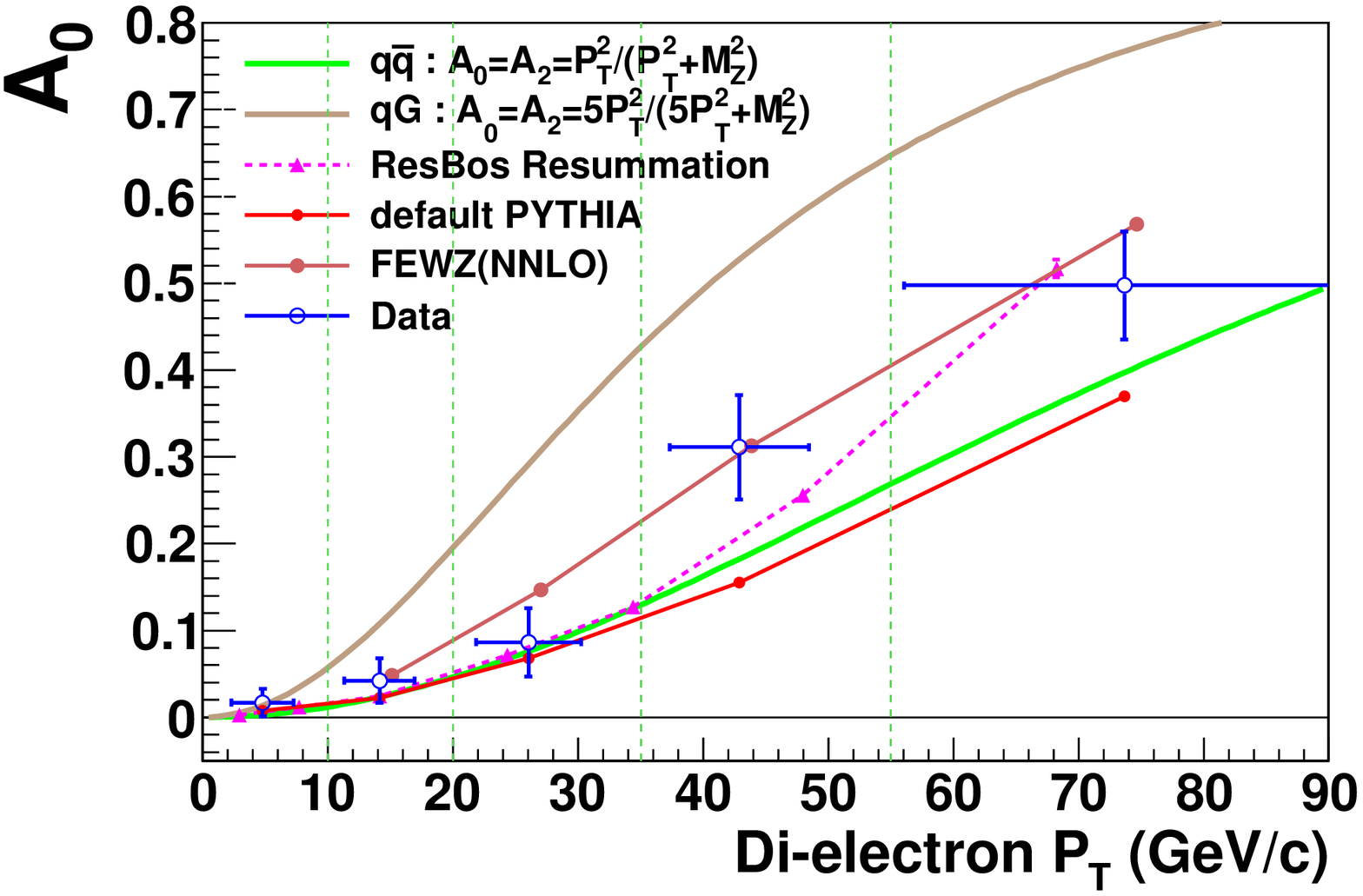}
  \includegraphics[width=80mm]{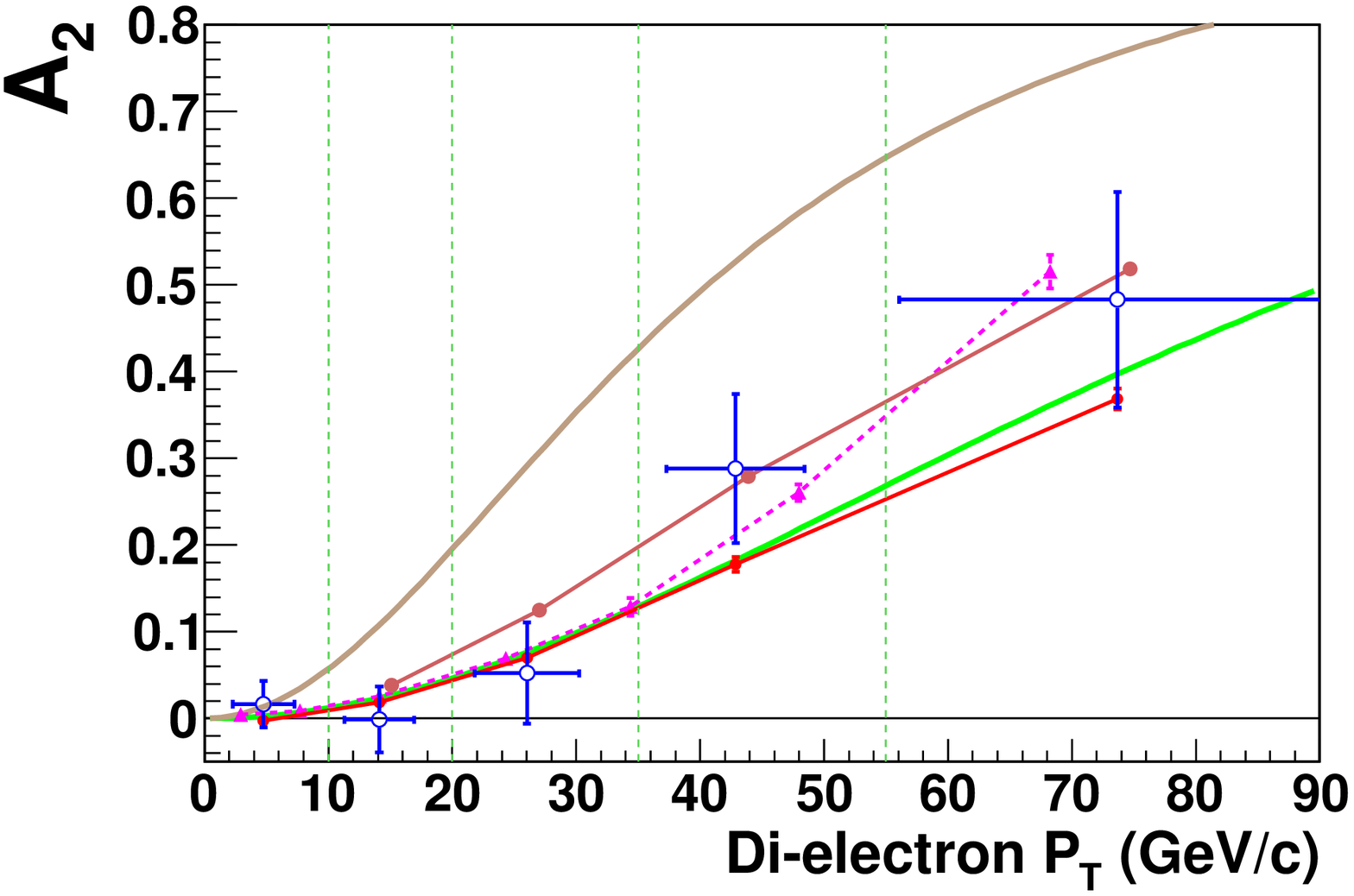}
  \caption{The measured $A_{0}$ and $A_{2}$ and predictions in $P_{\rm T}$
    for the mass window, $66<M_{ee}<116$ GeV/$c^{2}$. $A_{0}$ and $A_{2}$ have a strong $P_{\rm T}$ dependence.
    the Lam-Tung relation ($A_{0}\simeq A_{2}$) which is only valid for vector gluon theory is confirmed. 
    (The average of $A_{0} - A_{2}$ is measured to be $0.017 \pm 0.023$.)
    For the measured $A_{0}$ and $A_{2}$, the uncertainties correspond
    to the total uncertainties ($stat. \oplus syst.$)} \label{fig:result1}
\end{figure}

\begin{figure}[ht]
  \centering
  \includegraphics[width=80mm]{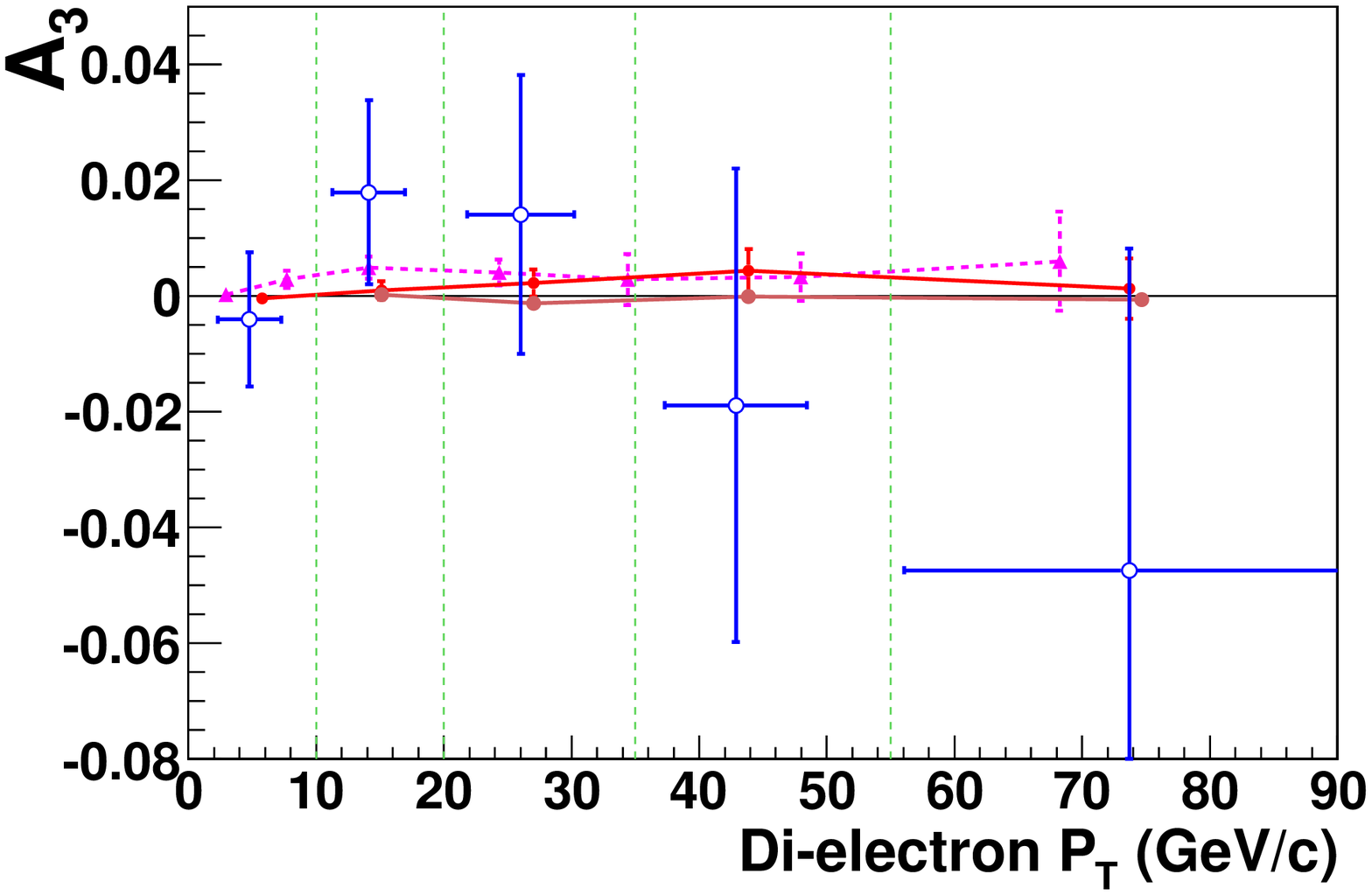}
  \includegraphics[width=80mm]{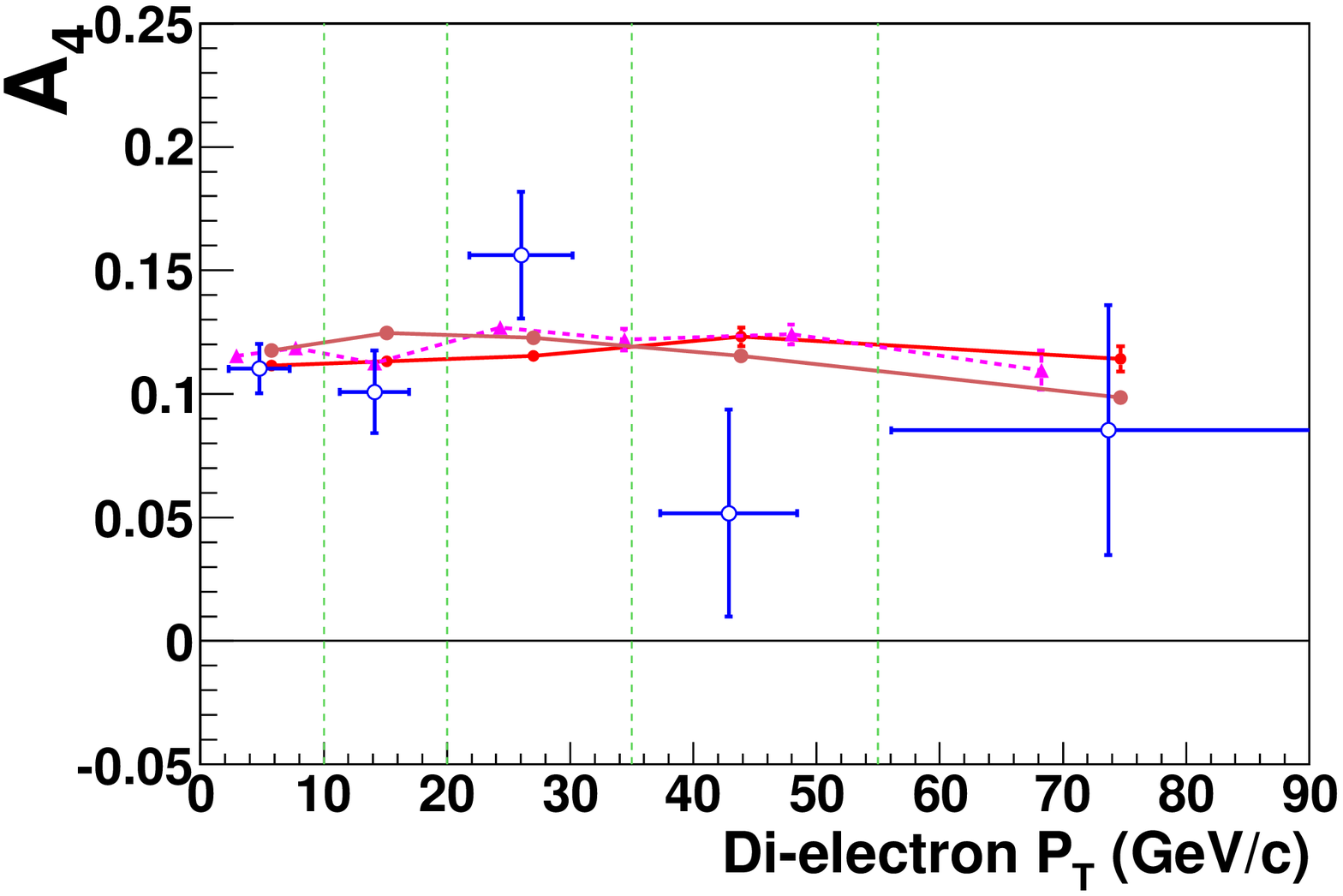}
  \caption{The measured $A_{3}$ and $A_{4}$ and predictions in $P_{\rm T}$
    for the mass window, $66<M_{ee}<116$ GeV/$c^{2}$. 
    $A_{3}$ and $A_{4}$ are relatively flat in $P_{\rm T}$.
    For the measured $A_{3}$ and $A_{4}$, the uncertainties correspond
    to the total uncertainties ($stat. \oplus syst.$)} \label{fig:result2}
\end{figure}

The angular coefficients, $A_0$, $A_2$, $A_3$, and $A_4$, are extracted from the data in the mass region,
$66<M_{ee}<116$ GeV/$c^{2}$ for $P_{\rm T}$ bins shown in  Figure \ref{fig:result1} and \ref{fig:result2}.
The measurement is compared with the various predictions which are \pythia~\cite{pythia}, \pythia +1jet \cite{pythia64}, 
QCD calculations in the fixed order perturbation theory ({\sc dyrad}~\cite{dyrad}, {\sc madgraph}~\cite{madgraph},  {\sc powheg}~\cite{powheg}, and {\sc fewz})~\cite{FEWZ},
and QCD calculations with resummation ({\sc resbos}~\cite{resbos} and {\sc vbp}~\cite{vbp}).
The {\sc pythia}~and {\sc vbp}~predictions are close in $P_{\rm T}$ and follow the approximation of $q\bar{q}$
annihilation process, $\frac{P_{T}^{2}}{P_{T}^{2} + M_{\ell \ell}^{2}}$.
Other higher order predictions like {\sc dyrad}, {\sc madgraph},  {\sc powheg}, {\sc fewz}, and {\sc pythia}+1jet
are close and have the higher $A_{0}$ and $A_{2}$ values than {\sc pythia}~in high Z $P_{\rm T}$ region.
The {\sc resbos}~prediction follows {\sc pythia}~in low $P_{\rm T}$ region, but cross over to the high 
order predictions in $P_{\rm T} > 35$ GeV.
\par
$A_{0}$ and $A_{2}$ have a strong $P_{\rm T}$ dependence.
The measured $A_{0}$ and $A_{2}$ follow {\sc pythia}~prediction in low $P_{\rm T}$ bin (up to $P_{\rm T}<35$),
but prefer the high order predictions ({\sc dyrad}, {\sc madgraph},  {\sc powheg},  {\sc fewz},
and {\sc pythia}+1jet ) in $P_{\rm T}>35$.
At low $P_{\rm T}$, the measured values of $A_0$ and $A_2$ are well discribed by the $q\bar{q} \to \gamma^{*}/Z g$ annihilation function 
$A_0 = A_2 = \frac{P_{T}^{2}}{P_{T}^{2} + M_{\ell \ell}^{2}}$.
At high $P_{\rm T}$, the larger values show that both the annihilation and Compton process contribute to the cross section.
Our results are in agreement with the fixed order perturbation theory calculations including 
{\sc dyrad}, {\sc madgraph},  {\sc powheg}, {\sc fewz}, and {\sc pythia}+1jet.
In addition, we confirm the Lam-Tung relation ($A_{0}\simeq A_{2}$) which is only valid for vector gluon theory.
The average of $A_{0} - A_{2}$ is measured to be $0.017 \pm 0.023$, which is consistent with zero within the uncertainty.
\par
The measured $A_{3}$ and $A_{4}$ parameters are relatively flat in Z $P_{\rm T}$.
The central value of $A_{3}$ parameter is slightly going down in $P_{\rm T}$, but consistent with zero within one deviation
of the uncertainty.
The $A_{4}$ parameter determines the forward and backward asymmetry of Drell-Yan process which has a mass dependence. 
Here, we measure $A_{4}$ integrating of the mass range, $66<M_{ee}<116 GeV/c^{2}$.
The $P_{\rm T}$ dependence of the average value of $A_4$ for the mass range,  $66<M_{ee}<116$ GeV/$c^{2}$, is in agreement 
with the predictions of all the models.
The $A_{4}$ parameter is sensitive to Weinberg angle, $\sin\theta_{W}^{2}$.
To compare the measured $A_{4}$ with the predictions, we generate all predictions setting $\sin\theta_{W}^{2}=0.232$.

\section{$A_{fb}$ Measurement}
\begin{figure}[ht]
  \centering
  \includegraphics[width=100mm]{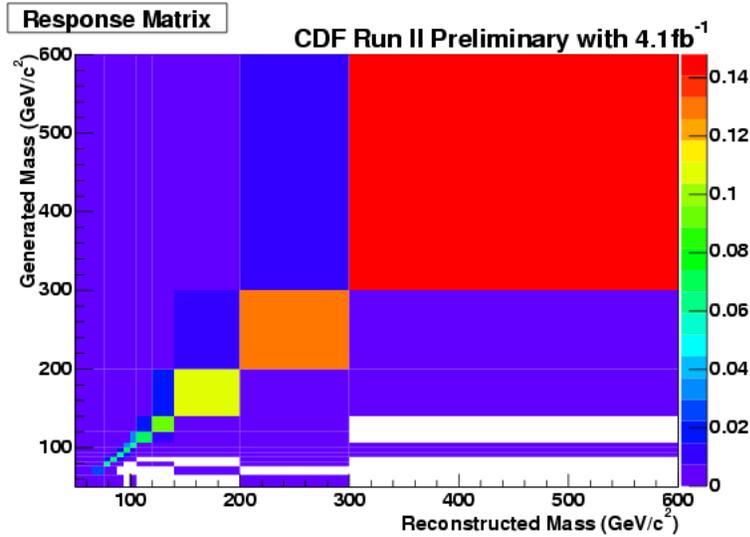}
  \caption{The response matrix for unfolding $A_{fb}$ measurement.
The response matrix includes the detector smearing effect, $A\times \epsilon$, and FSR effect. } \label{fig:response_matrix}
\end{figure}

\begin{figure}[ht]
   \centering
   \includegraphics[width=80mm]{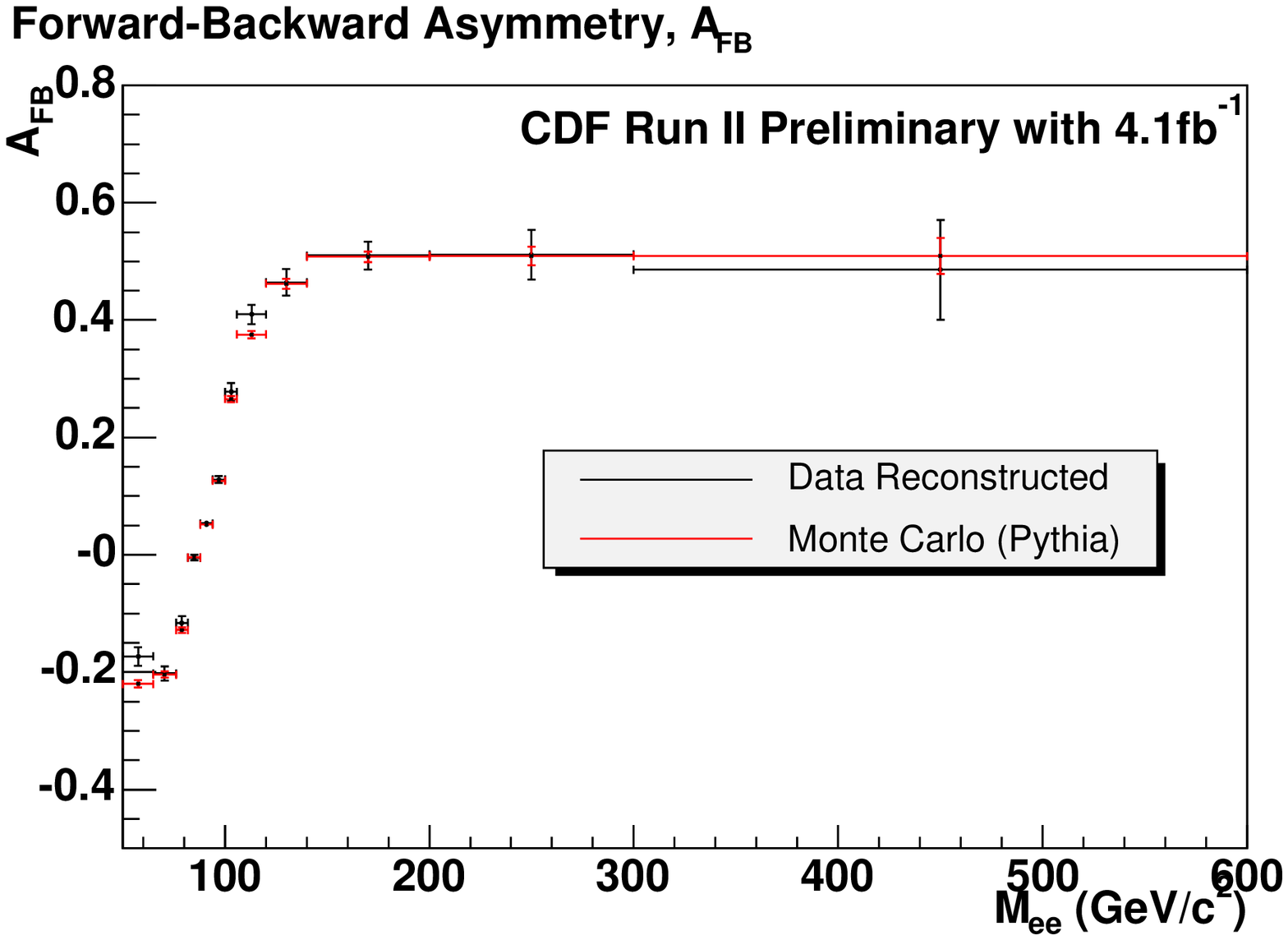}
   \includegraphics[width=80mm]{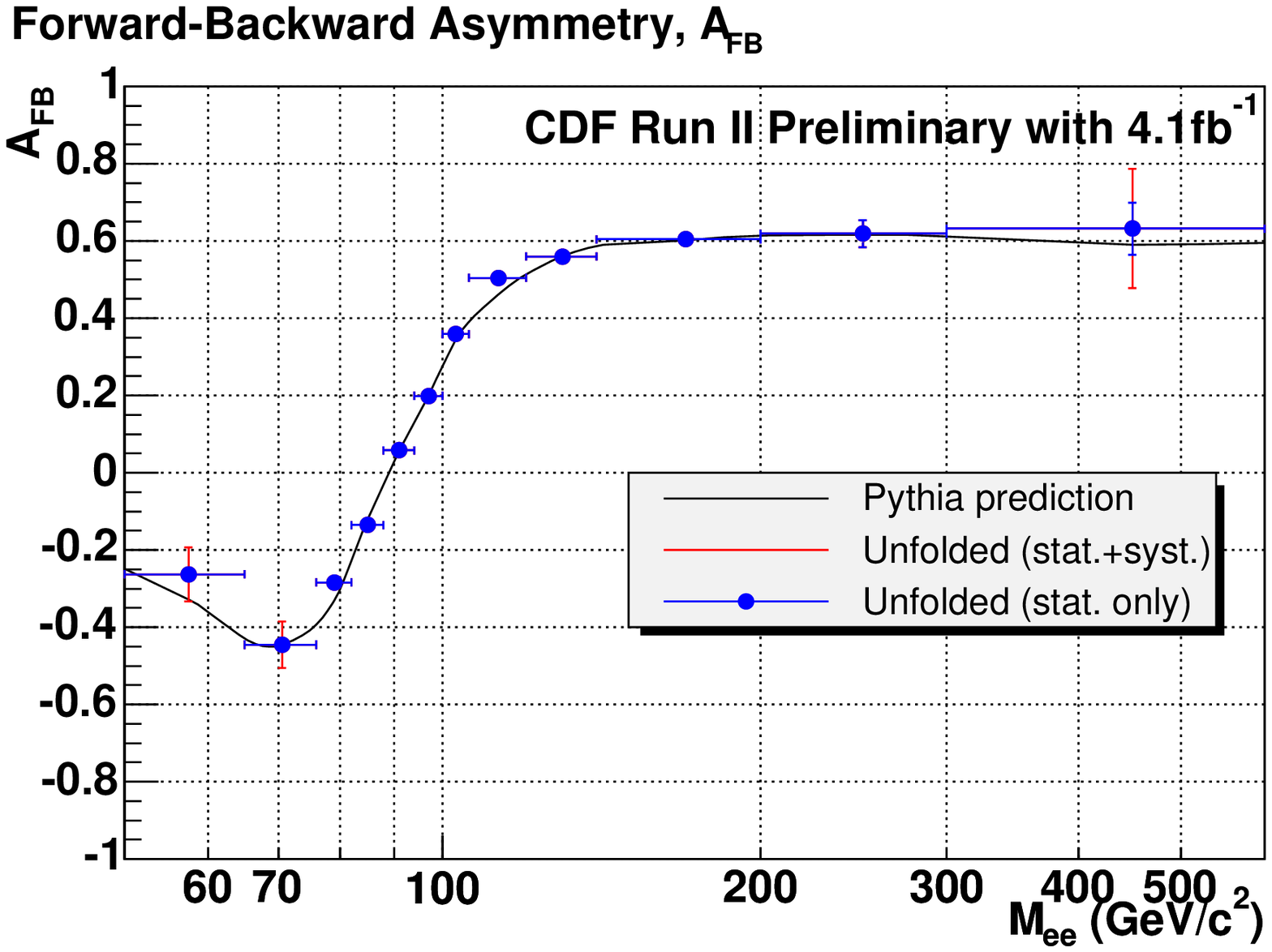}
   \caption{The measured $A_{fb}$ in dielectron mass before and after unfolding using 4.1 $fb^{-1}$ data.
     The left plot shows the measured $A_{fb}$ before unfolding and the right plot shows $A_{fb}$ after unfolding.
     For $A_{fb}$ before unfolding, the statistical uncertainty is only considered.
     For $A_{fb}$ after unfolding, the blue bars correspond to the statistical error 
     and the red bars to the total error (stat. $\oplus$ syst.).
     The $A_{fb}$ measurement is compared to \pythia~prediction. } \label{fig:afb}
 \end{figure}

$A_{fb}$ is measured using the event counting method in mass.
\begin{equation}
  \small
  A_{fb} = \frac{N_{sig}(\cos\theta > 0) - N_{sig}(\cos\theta < 0)}{N_{sig}(\cos\theta > 0) + N_{sig}(\cos\theta < 0)}
  \label{afb_definition}
\end{equation} 
 where $N_{sig}$ is the number of the signal events in the mass bin.
The measured $A_{fb}$ using 4.1 $fb^{-1}$ data is shown in Figure \ref{fig:afb} and it is compared with \pythia~prediction. 
This measured $A_{fb}$ is unfolded using the response matrix ($R_{ij}$) inversion method to get $A_{fb}$ in the physics level.
($\mu$ (true value) = $R_{ij}^{-1}(\nu$(observation) - $\beta$(background))
The response matrix includes the detector smearing effect, the acceptance and efficiency ($A\times \epsilon$) effect,
 and the final state photon radiation (FSR) effect. The response matrix is shown in  Figure \ref{fig:response_matrix}.
 
The measured $A_{fb}$ is unfolded using the response matrix inversion and the unfolded $A_{fb}$ is shown in Figure \ref{fig:afb}.
The unfolded $A_{fb}$ is compared to \pythia~prediction and it has a good agreement with the prediction.

\section{Weinberg angle ($\sin^{2}\theta_{W}$) vs. $A_{4}$}
\begin{figure}[ht]
   \centering
   \includegraphics[width=100mm]{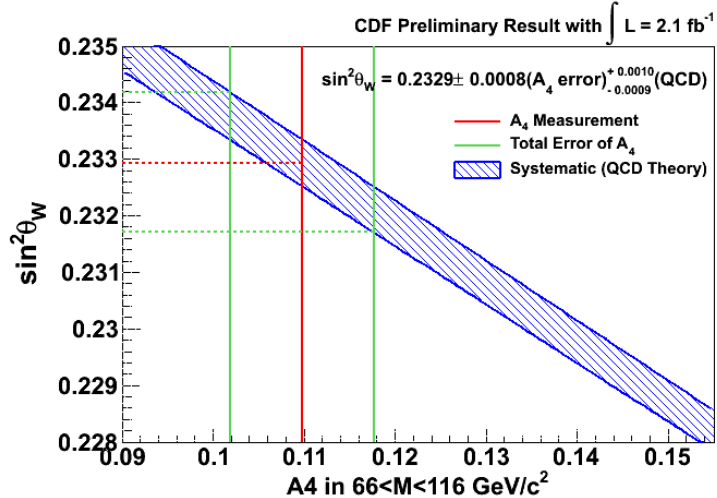}
   \caption{$A_{4}$ vs. $\sin^{2}\theta_{W}$ in $66<M_{ee}<116$ GeV/$c^{2}$.
     The theory band covers the order difference of the calculation and PDF difference (CTEQ vs. MSTW).} \label{fig:a4_mixing_angle}
 \end{figure}

One of the angular coefficients, $A_{4}$, is sensitive to the weak mixing angle, $\sin^{2}\theta_{W}$.
We translate $A_{4}$ measurement into $\sin^{2}\theta_{W}$ using the various predictions to extract the weak mixing angle
 in $66<M_{ee}<116$ GeV/$c^{2}$. The measured $A_{4}$ in \pt~is integrated over \pt~and $66<M_{ee}<116$ GeV/$c^{2}$.
For the predictions, \pythia, \resbos, \vbp, \powheg, and \fewz~are used to extract $\sin^{2}\theta_{W}$.
Figure \ref{fig:a4_mixing_angle} shows $A_{4}$ vs. $\sin^{2}\theta_{W}$ from various theory predictions.
The theory band in Figure \ref{fig:a4_mixing_angle} includes the order diffdrence of the calculation and also PDF difference
 (CTEQ vs. MSTW). The variation in the predictions is assigned as the uncertainty of QCD model.
The extracted $\sin^{2}\theta_{W} = 0.2329 \pm 0.0008$ ($A_{4}$ error)$^{+0.0010}_{-0.0009}$ (QCD).

\section{Conclusion}
We present the results of the first measurement of the angular coefficients in the production of $\gamma^{*}/Z$ bosons
at large transverse momenta, and the first test of the Lam-Tung relation at  high transverse momentum using 2.1 $fb^{-1}$ data.
We find a good agreement with the predictions of QCD fixed order perturbation theory, and with the Lam-Tung relation $A_0=A_2$ . 
From $A_{4}$ measurement, we extract Weinberg angle, $\sin^{2}\theta_{W}$, using various theory predictions. 
We also present $A_{fb}$ measurement in mass using 4.1 $fb^{-1}$ data and the measured $A_{fb}$ is compared to \pythia~prediction.
The measured $A_{fb}$ has a good agreement with \pythia~prediction.
We expect that a comparison of these result with the measurement at LHC would provide additional tests of production mechanisms
 because the contribution of Compton process at LHC is expected to be larger than Tevatron.
The angular measurement is published in Ref. \cite{angular_pub}.

\begin{acknowledgments}
We thank the Fermilab staff and the technical staffs of the participating institutions for their vital contributions. This work was supported by the U.S. Department of Energy and National Science Foundation; the Italian Istituto Nazionale di Fisica Nucleare; the Ministry of Education, Culture, Sports, Science and Technology of Japan; the Natural Sciences and Engineering Research Council of Canada; the National Science Council of the Republic of China; the Swiss National Science Foundation; the A.P. Sloan Foundation; the Bundesministerium f\"ur Bildung und Forschung, Germany; the Korean World Class University Program, the National Research Foundation of Korea; the Science and Technology Facilities Council and the Royal Society, UK; the Institut National de Physique Nucleaire et Physique des Particules/CNRS; the Russian Foundation for Basic Research; the Ministerio de Ciencia e Innovaci\'{o}n, and Programa Consolider-Ingenio 2010, Spain; the Slovak R\&D Agency; the Academy of Finland; and the Australian Research Council (ARC). 
\end{acknowledgments}

\bigskip 

\begin{thebibliography}{9}
\bibitem{cdf_coord}
  CDF coordinates are ($\theta$,$\phi$,$z$), where $\theta$ is the polar angle relative to
  the proton beam (the $+z$ axis), and $\phi$ the azimuth.
  The pseudorapidity is $\eta=-ln~tan(\theta/2)$. For an $e^{+}+e^{-}$ pair
  $P_{T}=Psin\theta$,  $E_{T}=Esin\theta$, 
  and $y=\frac{1}{2}ln\frac{E+P_{z}}{E-P_{z}}$, where
  $P$ and $P_{z}$ are the magnitude and z component of the momentum,
  and E is the energy of the $e^{+}+e^{-}$ pair.
\end{thebibliography}

\end{document}